\documentclass[nofootinbib,aps,groupedaddress,showpacs,floatfix,superscriptaddress,twocolumn]{revtex4-1}
\usepackage{amssymb}
\usepackage{amsmath}
\usepackage{graphicx}
\usepackage{color}

\newcommand{\bea}{\begin{eqnarray}}
\newcommand{\eea}{\end{eqnarray}}
\newcommand{\be}{\begin{equation}}
\newcommand{\ee}{\end{equation}}

\begin{document}

\title{Asymmetry relations and effective temperatures for biased Brownian gyrators}

\author{Sara Cerasoli}
\affiliation{Dipartimento di Fisica, Universit\`{a} degli Studi di Torino, via P. Giuria 1, 10125 Torino, Italy}
\author{Victor Dotsenko}
\affiliation{Sorbonne Universit\'e, CNRS, Laboratoire de Physique Th\'eorique de la Mati\`{e}re Condens\'ee, UMR CNRS 7600,
75252 Paris Cedex 05, France}
\affiliation{L.D.\ Landau Institute for Theoretical Physics,
           119334 Moscow, Russia}
\author{Gleb Oshanin}  
\affiliation{Sorbonne Universit\'e, CNRS, Laboratoire de Physique Th\'eorique de la Mati\`{e}re Condens\'ee, UMR CNRS 7600,
75252 Paris Cedex 05, France}
\author{Lamberto Rondoni}
\affiliation{Present address: Civil and Environmental Engineering Department and Princeton Environmental Institute,
Princeton University, 59 Olden St, Princeton, NJ 08540, USA}
\affiliation{Dipartimento di Scienze Matematiche, Dipartimento di Eccellenza 2018-2022,
Politecnico di Torino, Corso Duca degli Abruzzi 24, 10129 Torino, Italy}
\affiliation{INFN, Sezione di Torino, Via P. Giuria 1, 10125 Torino, Italy}
\affiliation{Malaysia-Italy Centre of Excellence for Mathematical Sciences, Universiti Putra Malaysia, 43400 Seri Kembangan, Selangor, Malaysia
}

\date{\today}

\begin{abstract}
We focus on a paradigmatic 
two-dimensional model of a nanoscale heat engine, 
- the so-called Brownian gyrator - whose stochastic dynamics is described by a pair of coupled Langevin 
equations with different temperature noise terms. This model
is known to produce a curl-carrying non-equilibrium steady-state with persistent angular rotations. 
We generalize the original 
model introducing constant forces doing work on the gyrator,  for which
we derive exact asymmetry relations, that are reminiscent of the standard fluctuation relations.
Unlike  the latter, our relations concern instantaneous and not time averaged
values of the observables of interest.
We investigate the full two-dimensional dynamics as well as the dynamics 
projected on the $x$- and $y$-axes, so that information about the state of the system can 
be obtained from just a part of its degrees of freedom.
Such a state is characterized by effective ``temperatures" that can be measured in 
nanoscale devices, but do not have a thermodynamic nature.
Remarkably, the effective temperatures appearing in full dynamics are distinctly different 
from the ones emerging in its projections, confirming that they are not thermodynamic 
quantities, although they precisely characterize the state of the system.
\end{abstract}


\maketitle

While in the past statistical physics has been mainly devoted to the microscopic basis of 
the macroscopic behaviour, present day research is largely addressing ``small''
or non-thermodynamic systems, which either evolve spontaneously or are subjected to external 
drivings and constraints. Unlike macroscopic systems, that are by and large described by thermodynamics 
and linear response theory, these systems are still hard to be framed within a comprehensive theory. 

In fact, macroscopic observations amount to drastic projections from highly dimensional
spaces to spaces of observables that consist of just a few dimensions. This is the reason
why thermodynamics is so universal; those projections loose an enormous amount of
information about the microscopic dynamics, hence the properties of observables only
minimally depend on such microscopic details, provided a few conditions are met. Basically,
it suffices that atomic forces are short ranged and repulsive. Then, universality as well
as the equivalence of ensembles are established, and the resulting theory of macroscopic
objects very generally holds. On the contrary, the behavior of systems made of a
non-thermodynamic number of elementary constituents strongly depends on all defining
parameters, and a theory as widely applicable as thermodynamics can hardly be envisaged.

Nevertheless, one common facet of such systems is that their observables undergo 
non-negligible fluctuations. Therefore, the seminal paper \cite{ECM} 
represents a pioneering attempt towards a unified theory of 
fluctuating phenomena \cite{GGboltz,GC}.
Its chief result
is called Fluctuation Relation (FR), and it constitutes one of the 
first exact results obtained for systems which are almost arbitrarily far from equilibrium. Close to equilibrium
the FR reproduces the Green-Kubo and Onsager relations \cite{GGons,ESR05}. 
Also, transient relations provide a method to investigate equilibrium properties of given systems, by 
means of non-equilibrium experiments, closing the circle with the Fluctuation Dissipation Relation, that 
yields non-equilibrium properties by means of equilibrium experiments. 
Various derivations have been given for such an asymmetry relation, that is informally written as:
\begin{equation}
\frac{{\rm Prob}(\sigma_\tau \approx A)}{{\rm Prob}(\sigma_\tau \approx -A)} \approx  e^{\beta_{\rm eff}A} ~, \quad
\sigma_\tau = \int_0^\tau {\bf J} \cdot {\bf F} ~ {\rm d} t
\label{firstFR}
\end{equation}
where $\sigma_\tau$ is the power dissipated in a long time interval $\tau$, for a system 
driven by a force {\bf F} that produces the fluctuating current {\bf J}, and 
${\rm Prob}(\sigma_\tau\approx A)$ is the probability that $\sigma_\tau$ is close to $A$.
For homogeneous systems at temperature $T$, measured in units of the Boltzmann constant $k_B$, 
 $\beta_{\rm eff}$ equals $1/T$ but in general it contains an ``effective temperature'' 
\cite{leti,sar,Cilibert2016} that depends
on the case at hand.

Given the importance and success of such a FR \cite{SER2007,Cilib2009,Udo,CilibReview}, a wide variety 
of analogous results has been derived in quite different contexts. In particular, we now have FRs for 
dynamical systems and stochastic processes, for classical and quantum systems, 
for transient, steady states and ageing systems, for global and local quantities, for steady 
and time dependent drivings. Such FRs concern 
a variety of observables, the most common involving work, heat and energy dissipation. In particular,
steady state FRs  for thermodynamic quantities need $\tau$ large compared to the microscopic 
dynamics time scales.
Although fluctuations are not 
normally observable in macroscopic systems, FRs have been verified in gravitational wave 
detectors, that are indeed meant to reveal microscopic fluctuations in macroscopic systems \cite{AURIGA}.
The literature on FRs is abundant 
\cite{RonMej,BPRV,GGnoneqbook,Kurchan,Pillet,EvSeAdvances,Udo,CilibReview,SEWR16,CengR,Pol}.

Despite so much activity,  it  still remains to clarify the extent to which symmetries 
analogous to eq. \eqref{firstFR} hold for non-thermodynamics or even non-physical 
phenomena, such as population dynamics, when described by models analogous to the physical ones.
However, the thermodynamic interpretation will not apply, in general.

In this respect, exactly solvable models are most useful, because they  
provide meaningful benchmarks for a more general analysis, unveiling at the same time 
the complexity and non-trivial aspects of the problem. 
To this end, we focus on the exactly solvable 
model of a Brownian gyrator widely studied in the past (see below), due its non-trivial non-equilibrium steady-state (NESS)
with a non-zero curl. We generalize this model here introducing constant forces that do
work on the gyrator.
This does not merely add another variable to the parameter space, but 
allows us to go well beyond the previous analyses and probe asymmetry relations, reminiscent of
the standard FRs, for such a curl-carrying NESS. 
We derive exact asymmetry relations, characterizing the state of the system, 
for the two-dimensional as well as for the one-dimensional projected dynamics, 
revealing a peculiar behavior for the corresponding effective ``temperatures". 
The functional form of the latter appears perhaps to be even more striking  
than the one obtained for the exchanged heat in transient and stationary regimes \cite{Cilibert2016}.
Moreover, our relations are more detailed than standard ones, because they concern
instantaneous values, rather than averaged values of the observables of interest.
 
Stochastic dynamics of a \textit{biased} 
Brownian 
gyrator is described by two coupled 
Langevin equations:
\begin{align}
\label{1}
\dot{x} =  F_x - \frac{\partial U_0}{\partial x}(x,y) + \zeta_x(t) \,, \nonumber\\
\dot{y} = F_y - \frac{\partial U_0}{\partial y}(x,y) + \zeta_y(t) \,,
\end{align}
in which 
the 
viscosity $\eta = 1$,  ${\bf F} = (F_x,F_y)$ is a constant force exerting a regular \textit{bias} on the 
gyrator (both $F_x$ and $F_y$ can independently take any real value), and the potential $U_0(x,y)$
has a generic parabolic form
\begin{align}
\label{pot}
U_0(x,y) =  \frac{x^2}{2} + \frac{y^2}{2} + u x y \,,
\end{align}
where the coupling constant $u$ obeys 
$u^2 < 1$, for reasons to be clarified below. 
Lastly, $\zeta_{\alpha}(t)$, ($\alpha = x,y$), are Gaussian white noises with zero mean and 
covariances  functions:
\begin{align}
\label{noises}
\overline{\zeta_{\alpha}(t) \zeta_{\beta}(t')} = 2 T_{\alpha} \delta_{\alpha,\beta} \delta(t - t') \,, 
\end{align} 
with $
\delta_{\alpha,\beta}=0$ for $\alpha \ne \beta$, and $\delta_{\alpha,\beta} = 1$ for $\alpha=\beta$. In general, $T_x \neq T_y$.
Note as well that
 the minimum $O_m =(x_m,y_m)$ of the 
effective potential $U(x,y) = - F_x x - F_y y + U_0(x,y)$, in which  the Langevin dynamics
takes place, 
is located at
\begin{align}
\label{center}
x_m = \frac{F_x - u F_y}{1 - u^2} \,, \,\, y_m = \frac{F_y - u F_x}{1 - u^2} \,,
\end{align}
i.e., it does not lie in the origin $(0,0)$, unless $F_x = 0$ and $F_y = 0$. For $u \to 1^-$, both $|x_m|$ and $|y_m|$ 
tend to infinity. 

Before we proceed, we note that the system described by eq. \eqref{1} can be viewed from a different perspective, for which
the non-thermodynamic character of the effective temperatures derived below is most apparent. Namely, the Langevin equations  \eqref{1} 
can be thought of as two rate equations describing the temporal evolution of, say, the "densities" $x$ and $y$ 
of two interacting "populations" which are continuously introduced into the system by two independent random sources with mean intensities $F_x$ and  $F_y$ and fluctuations having \textit{different} amplitudes - $T_x$ and $T_y$. The species of the populations have their intrinsic equal life-times (terms $-x$ and $-y$ in the right-hand-side of eq. \eqref{1}) and compete effectively for some resources, such that (for $0 < u < 1$) an increase of the $x$- or $y$-population prompts a decrease of the $y$- or $x$-population (terms $- u y$ and $- u x$ in the right-hand-side of eq. \eqref{1}). As we set out to show in what follows, dynamics of such coupled populations in this seemingly simple model appears to be rather non-trivial when the amplitudes of noises are not equal to each other. We note, as well, that such a rate equations approach will produce a physically plausible behavior only in some range of intensities $F_x$ and $F_y$. Outside of this range some spurious effects will take place, e.g., the densities will become negative.

Reference \cite{pel}, which addressed the notion of {\em effective temperatures} \cite{CKP},
first noted that the \textit{unbiased} case, $F_x = F_y = 0$, can be solved exactly in the $t \to \infty$ limit.
In Ref.\cite{1}, the model was seen as the simplest nanoscale heat engine; its
average torque was determined analytically,
and experimental realizations were discussed, including devices with: 
(a) an anisotropic black-body radiation; (b) an electrical heat bath made of two resistors at different 
temperatures; (c) two heat baths - a usual fluid environment with isotropic properties and an 
unusual one emitting thermal fluctuations in a preferential direction. 
In Refs.\cite{al,al1}, an experimental scenario close to (b) was realized, and the exchanged heat and work 
were measured. The theoretical analysis of Refs.\cite{al,al1}, consistent with the experimental evidence,
was based on a variation of eqs.\eqref{1},\eqref{noises} and \eqref{pot} suitable for the evolution of 
voltages in coupled resistors, produced fluctuation relations for the unbiased case. 
In Ref.\cite{Cilibert2016} this analysis was extended to obtain both transient and steady
states fluctuation relations, that respectively hold for all and only for asymptotic observation time
intervals, in accord with our discussion above. An experimental realization of gyrators was
developed also in Ref.\cite{Argun}, in a framework that can be generalized to include our model settings; that being, 
constant bias exerted on the gyrator.

Reference \cite{crisanti} used eqs.\eqref{1},\eqref{noises} and \eqref{pot} with $F_x=F_y=0$ to investigate 
the relevance of information contained in cross correlations among different degrees of freedom in 
non-equilibrium systems.
In turn,
Ref.\cite{2} and later Ref.\cite{3} focussed on non-equilibrium currents and provided explicit expressions 
for their curl, for the mean angular velocity of the rotational motion \cite{2,3}, and
for the variance of the latter \cite{3}. Reference \cite{3} also argued that eqs.\eqref{1}, \eqref{noises} and 
\eqref{pot} may describe a rotation of clouds of cold atoms, following laser detuning imbalance 
during the cooling phase, which leads to different temperatures along the different cooling axis.
Lastly, Refs.\cite{2,3} numerically analyzed the time-averaged angular velocity $\omega$ for 
a single long trajectory $\rho_t$, demonstrating that $\omega$ converges 
to its ensemble-averaged counterpart in the limit of a long observation time. 

Our model with constant forces doing work on the gyrator  generalizes the previous ones, so that more experimental 
devices, including {\em e.g.}\ nanomechanical resonators \cite{Tamayo,Riccia},  and even
population dynamics , can be considered \cite{inpreparat}.
Like other investigations of fluctuation relations led to results of more general interest
({\em e.g.}\ novel response theories \cite{Ruelle,SEWR16}), our asymptotic 
{\em asymmetry} relations produce a new form of effective temperature, expected to be measurable
{\em e.g.}\ in settings similar to those of Refs.\cite{al,al1,Riccia}.

\begin{figure*}
\begin{center}
\includegraphics[width=0.3\textwidth,height=0.3\textwidth]{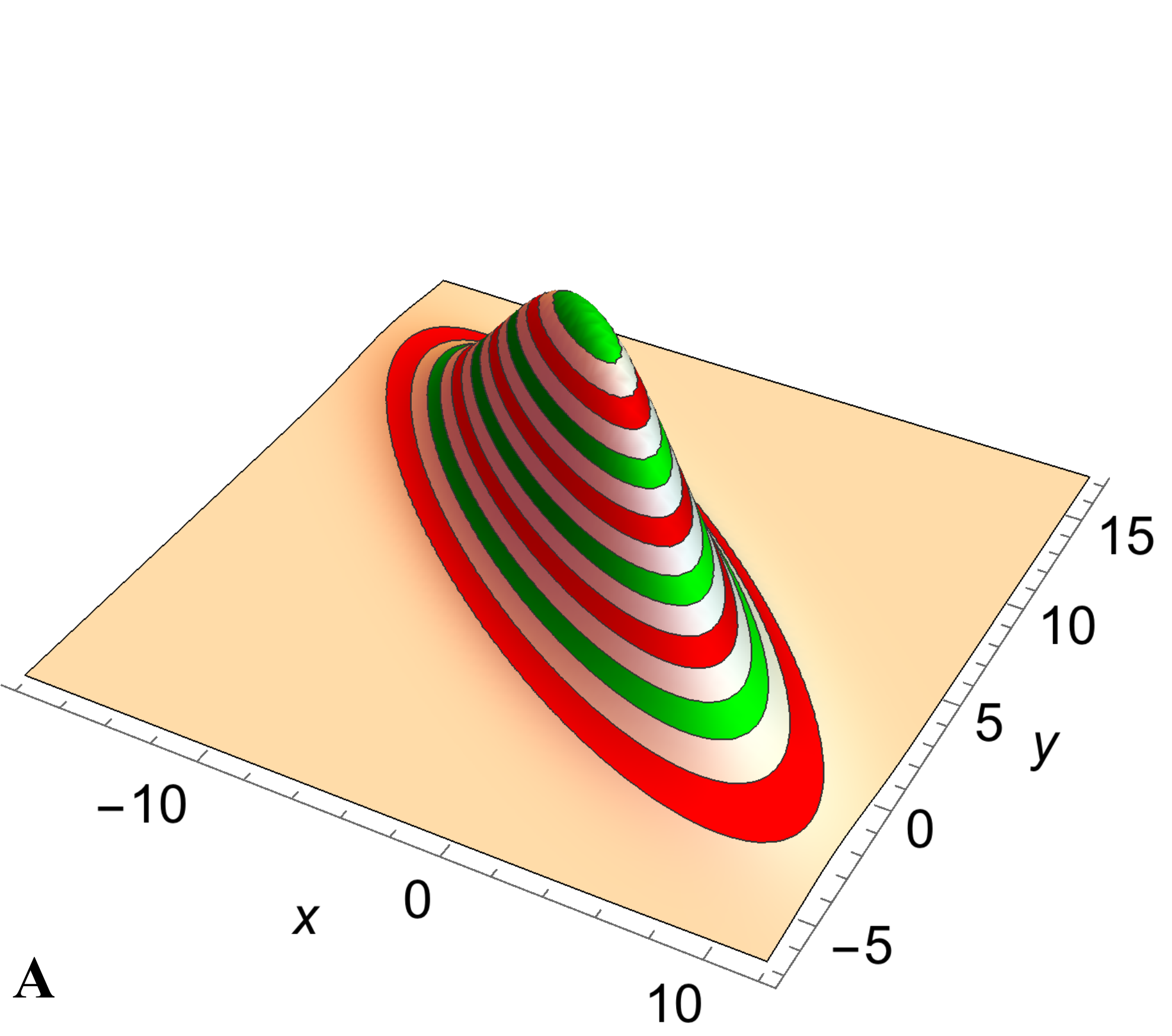}\hspace{10pt}
\includegraphics[width=0.25\textwidth,height=0.25\textwidth]{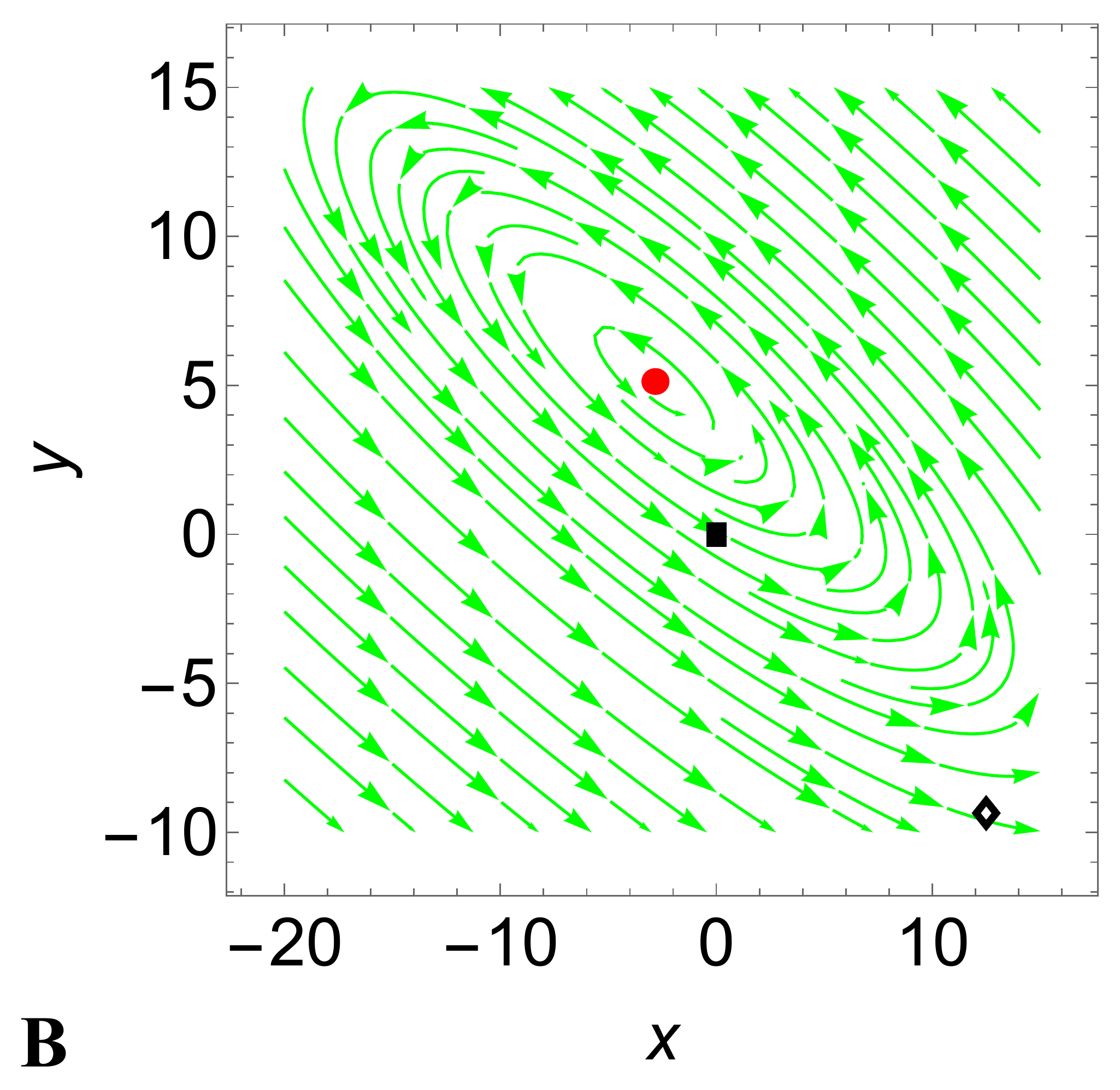}\hspace{10pt}
\includegraphics[width=0.3\textwidth,height=0.3\textwidth]{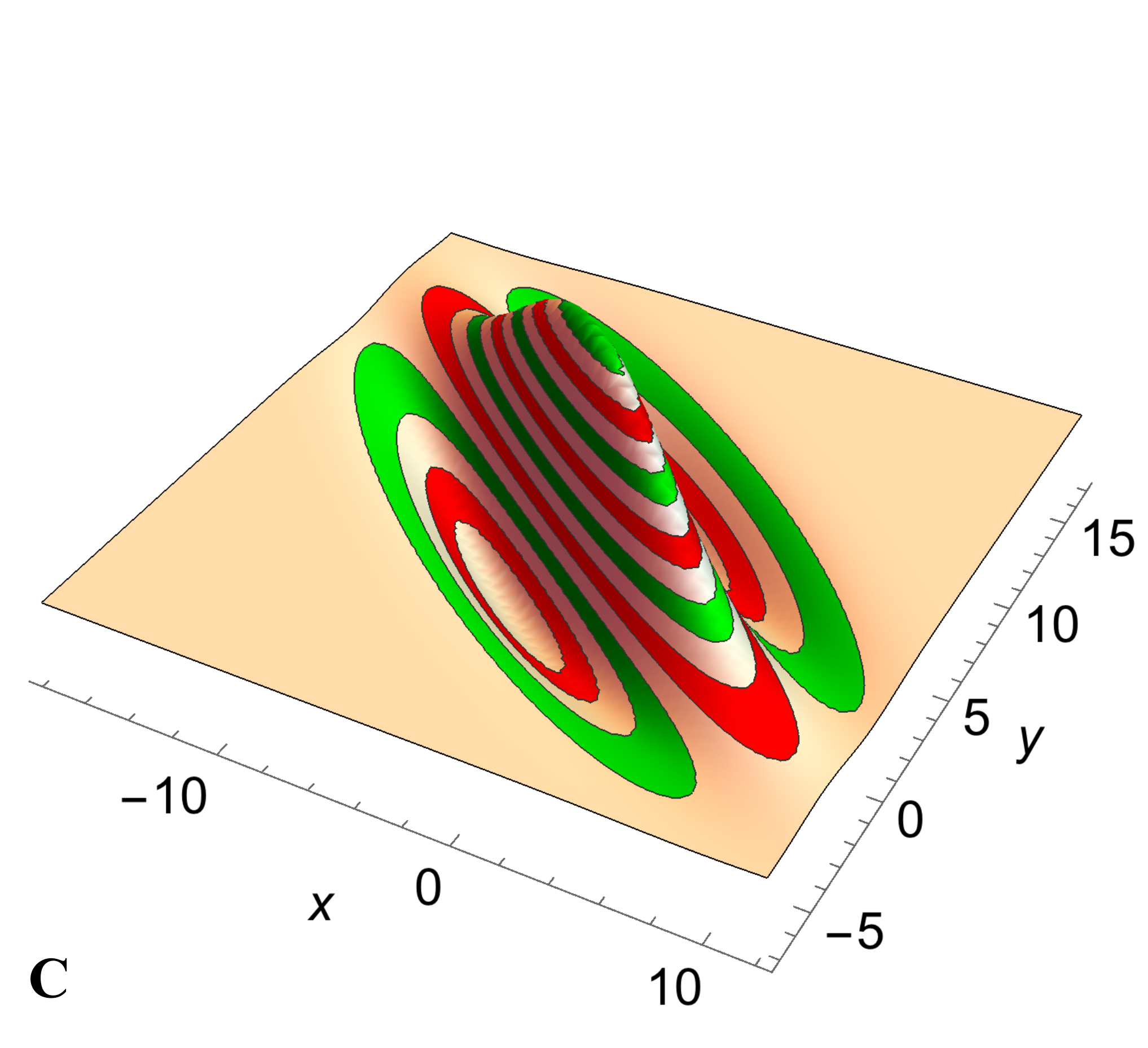}\hspace{10pt}
\includegraphics[width=0.3\textwidth,height=0.3\textwidth]{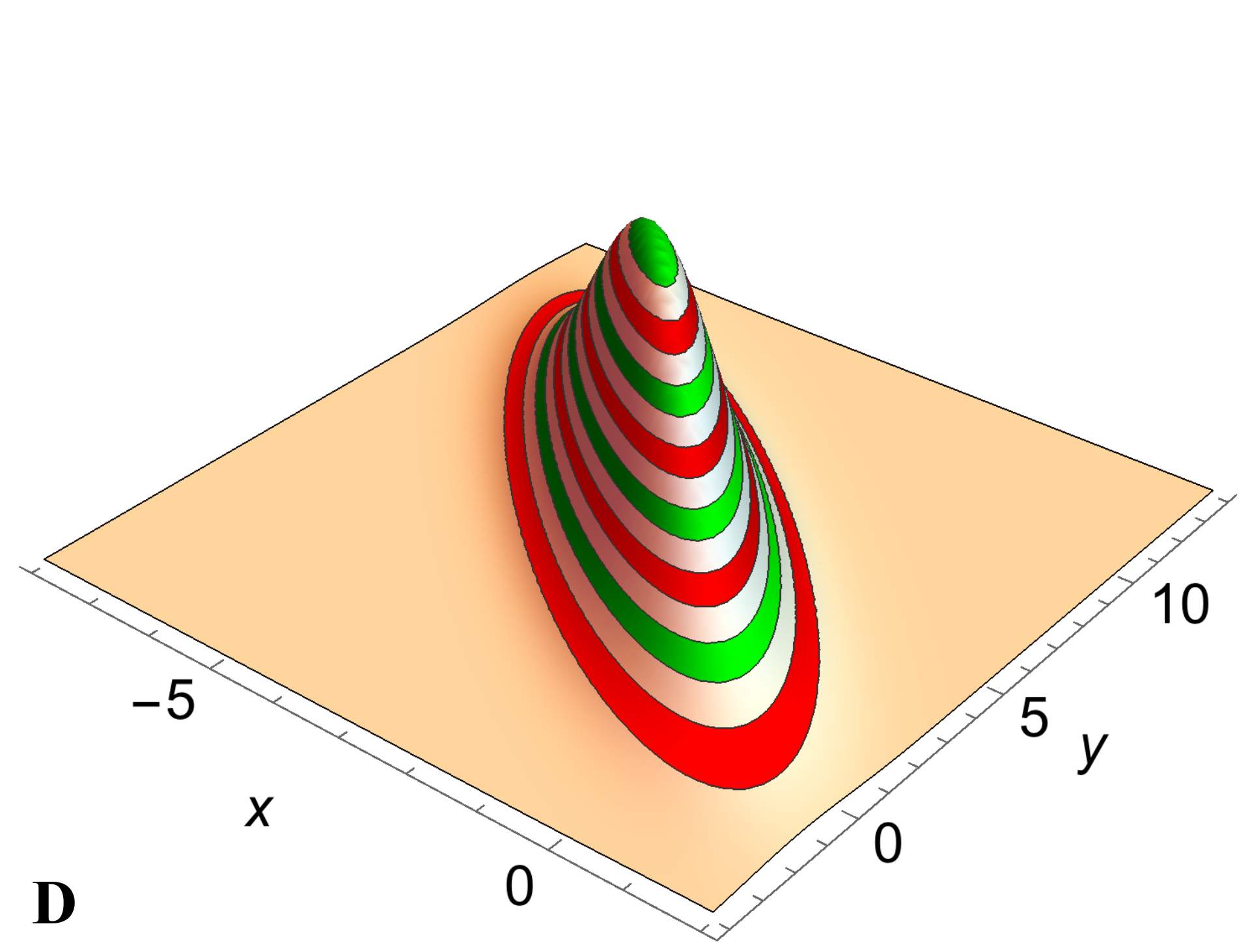}\hspace{14pt}
\includegraphics[width=0.25\textwidth,height=0.25\textwidth]{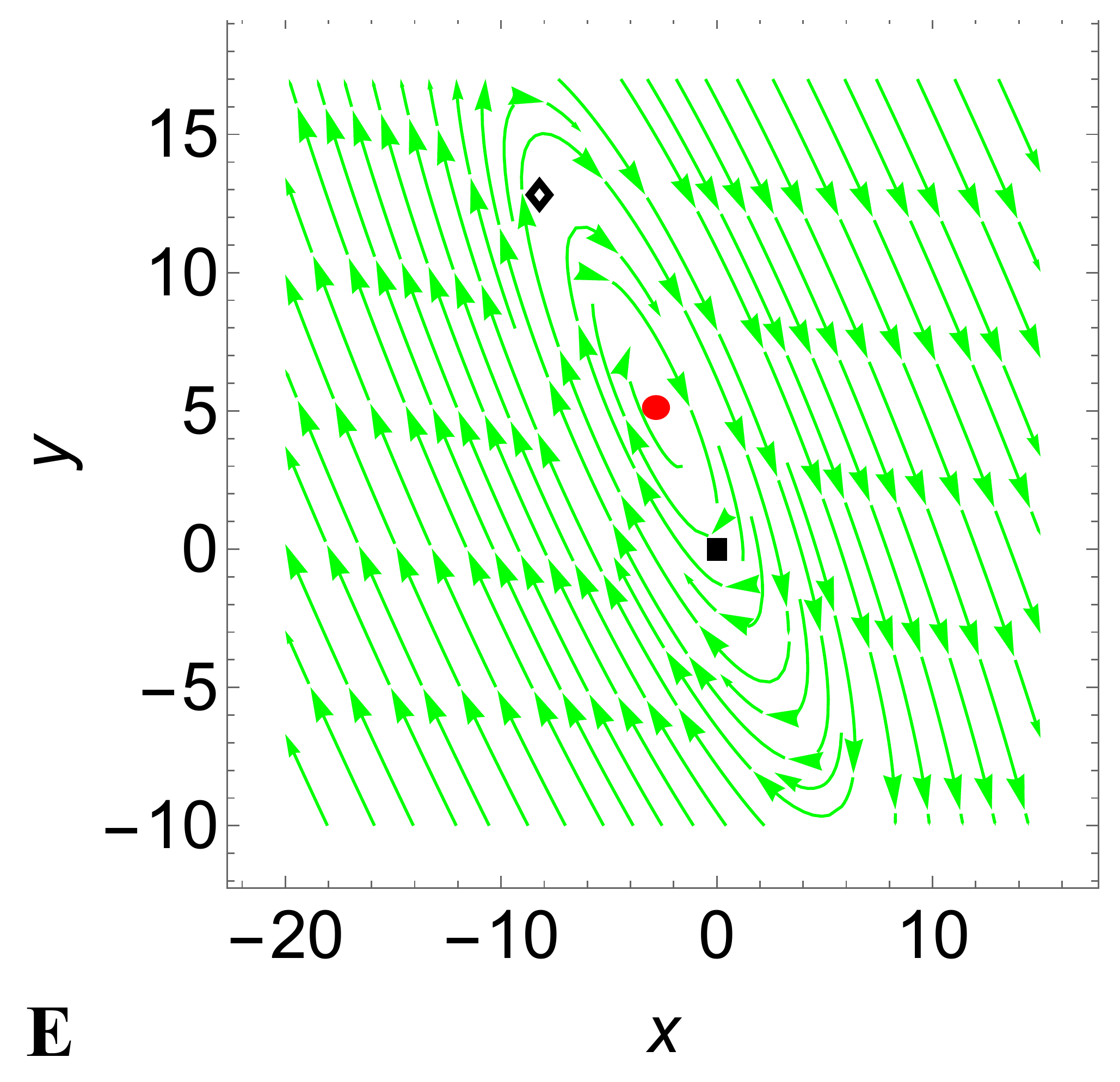}\hspace{10pt}
\includegraphics[width=0.3\textwidth,height=0.3\textwidth]{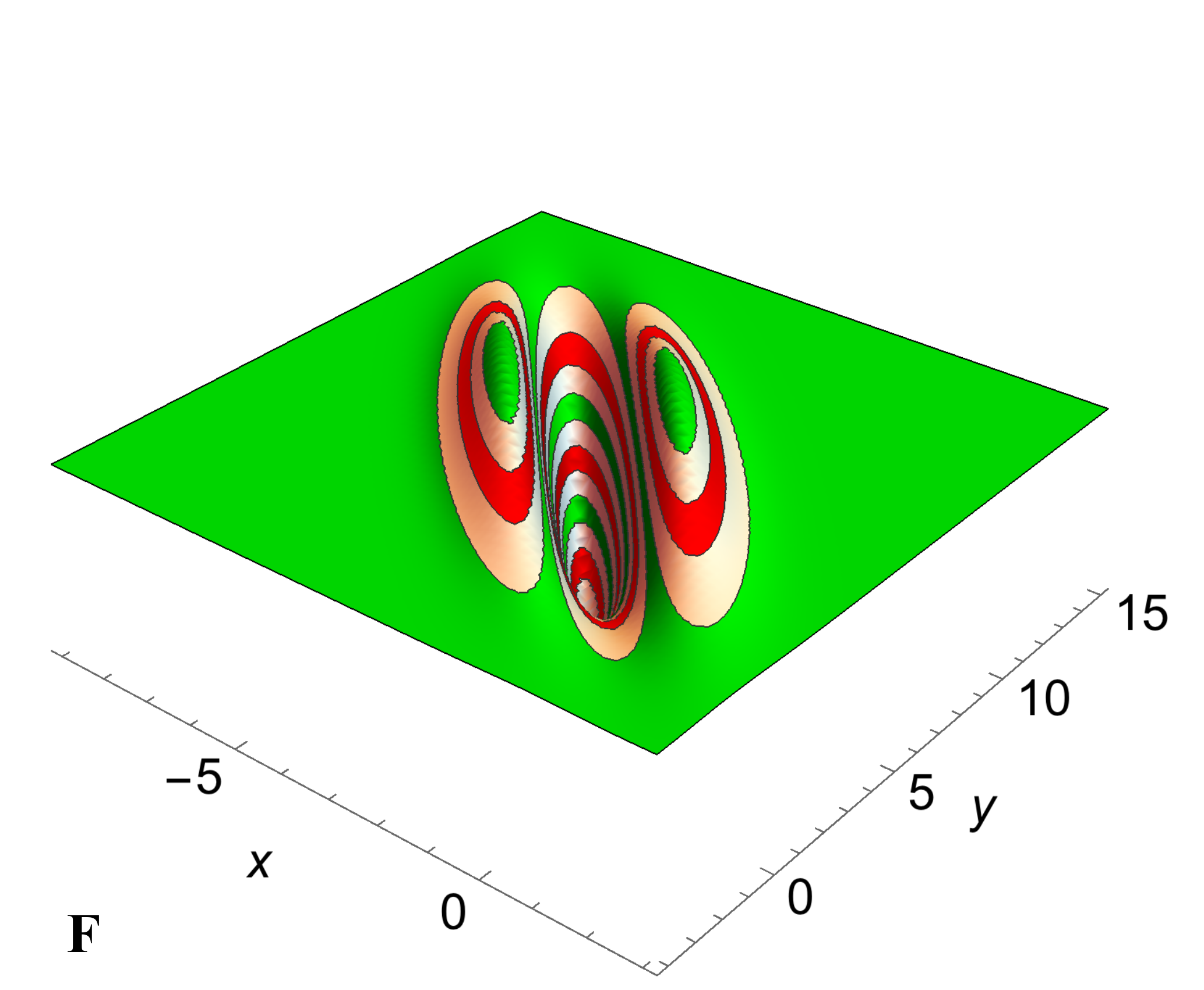}\hspace{10pt}
\caption{The joint pdf $P(x,y)$ (panels A and D), eq. \eqref{4}, the current ${\bf j}$ (panels B and E), eqs. \eqref{3}, and the curl of the current ${\bf j}$ (panels C and F), eq. \eqref{curl}, as functions of $x$ and $y$ for $u=3/4$ and for different choices of $T_x$, $T_y$, $F_x$ and $F_y$. In the first raw, $T_x = 20$, $T_y = 1$, $F_x = 1$ and $F_y = 3$. In the second raw, $T_x = 0$, $T_y = 5$, $F_x = 1$ and $F_y = 3$. In Panels B and E, the full square denotes the origin, the filled circle (red) denotes the minimum $O_m$ of the potential, eqs. \eqref{center}, while the diamond shows the location of point of isometry $O_{is}$ with coordinates $x_{is} = u \Delta_T (2 F_y T_x + F_x \left(T_x + T_y\right) - F_y u^2 \Delta_T)/(1-u^2) (4 T_x T_y + u^2 \Delta_T^2)$ and $y_{is} = - u \Delta_T (2 F_x T_y + F_y \left(T_x + T_y\right) + F_x u^2 \Delta_T)/(1-u^2)(4 T_x T_y + u^2 \Delta_T^2)$ (see the text above eq. \eqref{is}).}
\label{fig:fig2}
\end{center}
\end{figure*}

In the NESS, the Fokker-Planck equation for the probability density function $P(x,y)$ 
takes the form:
\begin{align}
\label{2}
{\rm div}({\bf j}) = \frac{\partial j_x}{\partial x} + \frac{\partial j_y}{\partial y}  = 0\,,
\end{align}
where the current ${\bf j}=(j_x,j_y)$ is defined by:
\begin{align}
\label{3}
j_x &=  T_x \frac{\partial P(x,y)}{\partial x}+ P(x,y) \frac{\partial U(x,y)}{\partial x} \,, \nonumber\\
j_y &= T_y \frac{\partial P(x,y)}{\partial x}+ P(x,y) \frac{\partial U(x,y)}{\partial x} \,.
\end{align}

The analytical solution of eqs. \eqref{2} and \eqref{3} reads: 
\begin{widetext}
\begin{align}
\label{4}
P(x,y)& = Z^{-1} \exp\Bigg(- \frac{2}{4 T_x T_y + u^2 \Delta_T^2} \Bigg(\left(T_y + \frac{u^2 \Delta_T}{2}\right) x^2 + \left(T_x - \frac{u^2 \Delta_T}{2}\right) y^2 + \nonumber\\
&+ u \left(T_x + T_y\right) x y -  \left(2 F_x T_y + u F_y \Delta_T\right) x -  \left(2 F_y T_x - u F_x \Delta_T\right) y\Bigg) \Bigg) \,, \\
\label{z}
Z &= \pi \sqrt{\frac{4 T_x T_y + u^2 \Delta_T^2}{1 - u^2}} \exp\left( \frac{2\left(F_y^2 T_x + F_x^2 T_y\right) - 2uF_x F_y \left(T_x+T_y\right) + u^2\left(F_x^2 - F_y^2\right) \Delta_T}{\left(1-u^2\right)\left(4 T_x T_y + u^2 \Delta_T^2\right)} \right) \,,
\end{align}
\end{widetext}
with $\Delta_T = T_x - T_y$. For $T_x = T_y = T$, one has the required Boltzmann form 
$P(x,y) = Z_0^{-1} \exp(- U(x,y)/T)$, while for $T_x \neq T_y$ and $F_x = F_y = 0$ the 
result reported in \cite{pel,1,al,al1,crisanti,2,3} is recovered.
Equation \eqref{z} shows that the normalization $Z$ exists only for $u^2 < 1$, as noted above. 

In Fig. \ref{fig:fig2} (panels A and D), we present the 
bivariate pdf $P(x,y)$ for $u=3/4$, and for two different choices of temperatures and forces. 
We observe that $P(x,y)$ has a maximum at $O_m$ and is strongly elongated along its principal axis, being 
considerably shorter in the perpendicular direction. The variances of the distribution relative to $O_m$ 
are given explicitly for all values of $T_x$ and $T_y$ by:
\begin{align}
\label{var}
\sigma_x^2 = \left \langle \left(x - x_0\right)^2 \right\rangle = \frac{1}{1 - u^2} \left(T_x - \frac{u^2}{2} \Delta_T \right)\,, \\
\sigma_y^2 = \left \langle \left(y - y_0\right)^2 \right\rangle = \frac{1}{1 - u^2} \left(T_y + \frac{u^2}{2} \Delta_T \right)\,,
\end{align} 
where the angular brackets, here and henceforth, denote averages with respect to the distribution of eqs.(\ref{4},\ref{z}).
Note: $\sigma_x^2$ and $\sigma_y^2$ are independent of $F_x$ and $F_y$, and 
they have opposite trends under variations of $\Delta_T$: 
for $\Delta_T > 0$ ($\Delta_T < 0$) $\sigma_x^2$ decreases (increases),
while $\sigma_y^2$ increases (decreases), for growing $| \Delta_T|$. 
When $u \to 1^-$ with $T_x$ and $T_y$ fixed, both variances grow without bounds.

In Fig.\ref{fig:fig2} (panels B and E) the vector plot of ${\bf j}(x,y)$ is given, showing 
that its circulation around $O_m$, along the closed orbits defined 
by $P(x,y) = {\rm const}$. The direction of such a circulation is determined by the sign of $\Delta_T$ only.

To further characterize {\bf j}, we compute its curl:
\begin{align}
\label{curl}
{\rm curl}({\bf j}) = \frac{\partial j_y}{\partial x} - \frac{\partial j_x}{\partial y} = \Delta_T u A(x,y) P(x,y) \,,
\end{align}
where $A(x,y)$ is a quadratic form of $x$ and $y$, with coefficients depending in a complicated fashion on the system parameters, which we omit here. Clearly, ${\rm curl}({\bf j})$ is rather non-trivial, as observed in Fig.\ref{fig:fig2} panels C and F. In particular, it qualitatively changes when $T_x$ and $T_y$ are changed: 
for $T_x = 20$ and $T_y = 1$ it has a hump and two deeps; while for $T_x = 0$ and $T_y = 5$ the hump becomes a deep, 
and the deeps turn into humps. This can be seen from the following expression for ${\rm curl}({\bf j})$ at $O_m$:
\begin{align}
\label{jj}
\left. {\rm curl}({\bf j})\right|_{x=x_m,y=y_m} = \frac{2 \sqrt{1-u^2} \left(T_x + T_y\right)}{\pi \sqrt{4 T_x T_y + u^2 \Delta_T^2}} \, u \Delta_T \,.
\end{align}
Here, $u \Delta_T > 0$ implies a maximum of the curl at the minimum of $U$, while if the inverse inequality 
holds, the minimum of $U$ implies minimal curl. Note that eq.\eqref{jj} does not depend
on the values of $F_x$ and $F_y$.

The ensemble-average angular velocity $\langle \omega \rangle$ is given by:
$\langle \omega \rangle = - \pi u \Delta_T/Z$, 
with $Z$ defined by eq.\eqref{z}.
In contrast to the variances, eqs.\eqref{var}, 
the curl and the rotation velocity strongly depend on ${\bf F}$. In particular, $Z$ brings an 
exponential dependence on $F_x$ and $F_y$ to $\omega$, which vanishes very rapidly when any
of the two forces increases. Indeed, the length of orbits increases 
with an increase of force, which also means a longer time to go round.  


Now, the pdf $P(x,y)$ obeys the asymmetry relation:
\begin{align}
\label{11}
\ln \frac{P(x,y)}{P(-x,-y)} & = \frac{2 F_x x}{T_{\rm eff}^{(x)}}+ \frac{2  F_y y}{T_{\rm eff}^{(y)}} +  \nonumber\\ 
 &+ u\Bigg( \frac{1}{T _{\rm eff}^{(y)}} -  \frac{1}{T _{\rm eff}^{(x)}} \Bigg)  \left(F_y x - F_x y \right)  \,.
\end{align}
Note that $F_x x$ and $F_y y$ are the works done by ${\bf F}$ along the directions $x$ ad $y$, for 
points starting at $(0,0)$, while $T_{\rm eff}^{(x)}$ and $T_{\rm eff}^{(y)}$ are effective temperatures, 
given  explicitly by:
\begin{align}
\label{9}
T_{\rm eff}^{(x)} = T_x + \frac{u^2}{4} \frac{\left(T_x - T_y\right)^2}{ T_y } \,,  \nonumber\\
T _{\rm eff}^{(y)}= T_y + \frac{u^2}{4} \frac{\left(T_x - T_y\right)^2}{T_x } \,. 
\end{align}
For reservoirs with $T_x = T_y = T$, both $T_{\rm eff}^{(x)}$ and $T_{\rm eff}^{(y)}$ 
equal $T$, while for $T_x \ne T_y$, the effective temperatures
are both \textit{larger} than  the thermodynamic temperatures of their respective reservoirs. Remarkably, a similar effect 
was experimentally observed and theoretically interpreted in Ref.\cite{EquipBr}.
Furthermore, letting the reservoir temperature $T_x$ vanish, so that the dynamics of $x$
is totally subordinated to that of $y$,  we observe that $T_{\rm eff}^{(x)}$ tends to $T_{\rm eff}^{(x)} = u^2 T_y/4$, 
while the effective temperature $T _{\rm eff}^{(y)}$ diverges. 
In contrast, the variances in eqs.\eqref{var} stay finite when $T_x \to 0$. Note, as well, 
that the divergence of $T _{\rm eff}^{(y)}$ derives from taking the $t \to \infty$ limit before 
the $T_x \to 0$ limit.
At finite times, the expression \eqref{11} should contain finite effective temperatures, with presumably 
exponentially fast divergence along $y$  with time. 
Finally, we can write:
\begin{align}
\label{10}
\frac{T_{\rm eff}^{(x)}+ T_{\rm eff}^{(y)}}{2} = \left(1 + \frac{u^2}{4} \frac{\left(T_x - T_y\right)^2}{ T_x T_y } \right) \frac{T_x + T_y}{2} \,,,
\end{align}
showing that the mean effective temperature is always greater than the mean temperature and, that it becomes 
infinitely large when either $T_x$ or $T_y$ vanish.

Equation \eqref{11} refers to the origin of the $(x,y)$-plane, but an analogous 
property can be obtained for any other point. Referring to $O_m$, one has: 
$\ln (P(x=x_m+\delta_x,y=y_m + \delta_y)/P(x=x_m-\delta_x,y=y_m - \delta_y)) \equiv 0$, for 
arbitrary $\delta_x$ and $\delta_y$. Also, there is a single point, $O_{is} = (x_{is},y_{is})$,
such that eq. \eqref{11} takes the isometric form 
\begin{align}
\label{is}
\ln \frac{P(x=x_{is} + \delta_x,y=y_{is} + \delta_y)}{P(x=x_{is} - \delta_x,y=y_{is} - \delta_y)} = \frac{2 F_x \delta_x}{T_{\rm eff}^{(x)}} +  \frac{2 F_y \delta_y}{T_{\rm eff}^{(y)}} \,,
\end{align}
for arbitrary $\delta_x$ and $\delta_y$, see Fig. \ref{fig:fig2} (panels B and E).

Marginalizing $P(x,y)$:
\begin{align}
P(x) &= \int^{\infty}_{-\infty} P(x,y) dy \,, \,\,\, P(y) = \int^{\infty}_{-\infty} P(x,y) dx \,.
\end{align}
one obtains the following pair of asymmetry relations:
\begin{align}
\label{41}
\ln \frac{P(x)}{P(-x)} &= \frac{2 x F_x}{\tau^{(x)}_{\rm eff} }  -2u \frac{F_y x}{\tau^{(x)}_{\rm eff} }   \,, \nonumber\\
\ln \frac{P(y)}{P(-y)} &= \frac{2 y F_y}{\tau^{(y)}_{\rm eff} }  -2u \frac{F_x y}{\tau^{(y)}_{\rm eff} }   \,,
\end{align}
with effective temperatures $\tau^{(x)}_{\rm eff} $ and $\tau^{(y)}_{\rm eff} $ given by
\begin{align}
\label{9}
\tau^{(x)}_{\rm eff} &= T_x - \frac{u^2}{2} \left(T_x - T_y\right) \,, \nonumber\\
\tau^{(y)}_{\rm eff}  &= T_y + \frac{u^2}{2} \left(T_x - T_y\right) \,. 
\end{align}
Unlike $T^{(x)}_{\rm eff}$ and $T^{(y)}_{\rm eff}$, $\tau^{(x)}_{\rm eff}$ and 
$\tau^{(y)}_{\rm eff}$ (a) can be smaller than $T_x$ and $T_y$; (b) do not 
diverge if either $T_x$ or $T_y$ vanishes; (c) are equal, up to a scale factor $1 - u^2$, to the
variances of the pdf $P(x,y)$, eq. \eqref{var}, and (d) they obey
 \begin{align}
\frac{\tau_{\rm eff}^{(x)}+ \tau_{\rm eff}^{(y)}}{2} \equiv \frac{T_x + T_y}{2} \,,
\end{align}
i.e., their mean  is identically equal to the mean reservoirs' temperature.  Interestingly enough, the effective temperatures emerging in the projected dynamics exhibit a completely different and somewhat trivial
behavior as compared to the one appearing in the full dynamics. This observation 
illustrates the statement made in the beginning of our work that projecting the dynamics
from higher
 dimensional spaces onto the individual components and marginalizing the distributions lead to a significant loss of information.

To conclude, we studied a model of a  Brownian gyrator subject to work-doing constant forces. 
We presented an exact solution of this model in the steady-state and showed that the 
probability density function  
obeys an asymmetry relation that contains effective temperatures, one of which may become 
arbitrarily large, mimicking certain experimental observations \cite{EquipBr}. 
The effective temperatures of 
the projected dynamics, on the contrary, cannot exceed the sum of the reservoirs temperatures.
Our results provide a novel important insight on the notion and behavior of effective temperatures 
in out-of-equilibrium conditions. In our case, they can be measured thanks to our 
relations, when the PDFs of the coordinates are known from {\em e.g.}\ experimental measurements,
such as those  performed in Ref.\cite{Argun}.  Lastly, we stress that the quantities discussed above may be interpreted physically, as
thermodynamic quantities, in particular. However, this interpretation is not necessary,
especially when small systems are observed over short
times. Furthermore, relations for
time-averaged quantities can be obtained by time integration, from our more detailed
instantaneous relations.

\section*{Acknowledgments}

SC wishes to thank LPTMC, Sorbonne Universit\'e, for a warm hospitality during her Master Degree training stage in July 2018 when 
this work has been performed, and also acknowledges support from Universit\`{a} degli Studi di Torino.
LR acknowledges partial support by MIUR grant Dipartimenti di Eccellenza 2018-2022.

\end{document}